\documentclass[twocolumn,english,xlinenumbers,aps,twocolumn]{revtex4-1}
\usepackage[T1]{fontenc}
\usepackage[utf8]{inputenc}
\usepackage{color}
\usepackage{babel}
\usepackage{amsmath}
\usepackage{amssymb}
\usepackage{graphicx}
\usepackage[unicode=true,pdfusetitle,
 bookmarks=true,bookmarksnumbered=false,bookmarksopen=false,
 breaklinks=false,pdfborder={0 0 0},pdfborderstyle={},backref=false,colorlinks=true]
 {hyperref}
\hypersetup{
 pdfborderstyle={},citecolor=blue}
 
\begin{document}
 \title{Impact of high-order effects on soliton explosions in the complex cubic–quintic Ginzburg–Landau equation}

\author{S. V. Gurevich$^{1,2}$}
\email{gurevics@uni-muenster.de}
\author{C. Schelte $^{1,3}$}
\author{J. Javaloyes$^{3}$}
\affiliation{$^{1}$Institute for Theoretical Physics, University of Münster, Wilhelm-Klemm-Str. 9, D-48149 Münster, Germany}
\affiliation{$^{2}$Center for Nonlinear Science (CeNoS), University of Münster, Corrensstrasse 2, D-48149 Münster, Germany}
\affiliation{$^{3}$Departament de Física, Universitat de les Illes Balears, C/ Valldemossa
km 7.5, 07122 Mallorca, Spain}

\begin{abstract}
We investigate the impact of higher-order nonlinear and dispersive effects on the dynamics of a single soliton solution in the complex cubic–quintic Ginzburg–Landau equation. Operating in the regime of soliton explosions, we show how the splitting of explosion modes is affected by the interplay of the high order effects (HOEs) resulting in the controllable selection of right- or left-side periodic explosions. In addition, we demonstrate that HOEs induce a series of pulsating instabilities, significantly reducing the stability region of the single soliton solution.
\end{abstract}

\maketitle

Soliton explosions are among the most striking and fascinating nonlinear phenomena studied in mode-locked lasers. They were first predicted theoretically in a one-dimensional complex cubic-quintic Ginzburg-Landau equation (CQGLE)~\cite{SCAA_PRL2000} for anomalous linear dispersion and then verified experimentally in a Kerr lens mode-locked Ti:sapphire laser~\cite{CSCA_PRL2002}. In this regime, a localized pulse circulating in the cavity experiences an abrupt structural collapse at certain points of its time evolution and subsequently recovers its original shape. Many numerical studies were reported in this framework~\cite{ASC_PLA2003,ASC_PRE2004,SCA_MCS2005,DB_PRE2010,DCJB_PRE2011,CDB_PRE2012,CD_PRE2013}. Among the reported features is the stable existence of symmetric and asymmetric explosive localized states (LSs) over a wide range of parameters. Further away from the explosion threshold, the exploding LSs experience a complex dynamics and exhibit spatio-temporal chaos following the Ruelle-Takens route; The LSs conserve an almost identical shape after each explosion cycle and the times between explosions appear to be randomly distributed. In two spatial dimensions, it was shown that the center of mass of asymmetrically exploding LS undergoes a subdiffusive continuous-time random walk despite the deterministic character of the underlying model, while in the case dominated by only asymmetric explosions, it becomes characterized by normal diffusion~\cite{CCDB_PRL12,CDAR_PRL16}.
Recently, exploding LSs have been observed in an all-normal-dispersion Yb-doped mode-locked fiber laser operated in a transition regime between stable and noise-like emission~\cite{RBE_Optica15}. The resulting experimental evidence has been successfully compared to realistic numerical simulations based on an envelope function approach~\cite{RAPEB_OFT2014}. There, the observed explosions manifest themselves as abrupt temporal shifts in the output pulse train. In~\cite{CD_PRA2016}, the connection between the pulse propagation model in optical fibers developed in \cite{RAPEB_OFT2014} and the CQGLE with additional higher-order nonlinear and dispersive effects was established. The latter leads to a formation of periodic, non-chaotic \emph{one-side} explosions. Further, it was shown that the mechanism of the explosion formation is different from the CQGLE without higher-order effects (HOEs): the periodic explosions result from a period-halving followed by a period-doubling bifurcation leading to the formation of chaotic explosions if the nonlinear gain parameter is varied. 

As was mentioned in~\cite{CSCA_PRL2002}, HOEs and in particular third-order dispersion can cause the asymmetry of the observed pulse explosion. The interactions between third-order dispersion and other higher-order effects become important for stable pulse generation~\cite{LLTZS_PRL2002,SLLZ_OC2005}. The influence of different HOEs on the exploding LSs in the CQGLE was studied numerically in~\cite{TLTZZ_APB2004, SLLZ_OC2005, LF_OL2010, FCF_PLA10,LF_OL11,LFF_APB2011, FC_PLA11, CF_PLA2012}. In particular, it was shown that a proper combination of the three HOEs can provide a shape stabilization of an exploding LS. However, despite significant theoretical interest, the impact of HOEs on the onset of soliton explosions have not been extensively studied so far and the full bifurcation study of the explosion regime is still lacking. 

In this paper we investigate the impact of three HOEs, namely, self-frequency shift, self-steepening, and third-order dispersion on the selection mechanism of soliton explosions in the CQGLE. Using path following techniques applied to partial differential equations (PDEs), we map the evolution of explosion regimes as a function of HOEs parameters and show how the interplay of the HOEs can result in non-trivial interaction of the explosion modes resulting in the selection of right- or left- periodic explosions for certain sets of system parameters. Finally we study the impact of HOEs terms on the stable LSs profile and we disclose new, HOEs induced, pulsating instabilities, leading to a significant reduction of the stability region of the single LSs.

The CQGLE is an amplitude equation describing the onset of an Andronov–Hopf bifurcation in dynamical systems~\cite{Newell}. In nonlinear optics, 
is considered as one of the paradigms for LS formation in mode-locked lasers and it is also widely used to describe such phenomena as short pulse propagation in optical transmission lines, dynamics of multimode lasers, parametric oscillators, and transverse pattern formation in nonlinear optical media~\cite{Weiss1993,Moores_OC93,3MIH_JLT95,SCP_PRE97,AAS_JOSAB98,GA_OE04,FS_PRL96,JTHTPL_OL99,AA-LNP-08}. In one dimension, the CQGLE with HOEs reads
\begin{eqnarray}\label{eq:CQGLE2}
 i\,\partial_zA&+&\frac{D}{2}\partial_{t}^2A+|A|^2\,A+\nu|A|^4\,A\\ \nonumber
 &=& i\delta A+i\epsilon|A|^2A+i\beta \partial_{t}^2A+i\mu|A|^4\,A+\mathrm{H.O.E.}\,,
\end{eqnarray}
Here, $z$ is the normalized propagation distance (or the cavity round-trip number when used to describe passively mode-locked lasers), $t$ is the retarded time (or a transversal spatial coordinate), $A$ is the normalized envelope of the field, $D=\pm 1$ is the group velocity dispersion coefficient corresponding to a anomalous or normal regime, $\delta>0$ ($<0$) is the linear gain (loss) coefficient, $\beta>0$ accounts for spectral filtering,  $\mu<0$ represents the saturation of the nonlinear gain, $\nu$ corresponds to the saturation of the nonlinear refractive index and $\epsilon$ is the nonlinear gain parameter. While the classical cubic CGLE describes a supercritical bifurcation, in the case of subcritical instability this equation is augmented with a fifth-order terms to allow the existence of stable pulse-like localized solutions if $\delta<0$ and $\epsilon>0$. The HOE contributions are given by the nonlinear gradient and third order dispersion terms:
$$
\mathrm{H.O.E.}=i\beta_3\partial_{t}^3A-is\partial_t(|A|^2\,A)+\tau_R\,A\partial_t|A|^2\,.
$$
where $\beta_3$ accounts for the third order dispersion (TOD), whereas the last two terms represent the
nonlinear gradient terms. Here, $s$ corresponds to the self-stepping (SST) and $\tau_R$ is a coefficient related to the intrapulse Raman scattering (IRS), which determines the soliton self-frequency shift. Note that in an envelope expansion near onset, the nonlinear gradient terms occur to the same order as the quintic term and cause the fixed-shape solution to be asymmetric and to move at a velocity other than the group velocity~\cite{DB_PLA1990}. Note also that in the pulse propagation model in optical fibers \cite{RAPEB_OFT2014} a non-local term accounting for both instantaneous electronic and delayed Raman contributions is considered. However, in \cite{CD_PRA2016} it was shown that for pulses that are wide enough ($\sim$ 0.1 ps), the general non-local pulse propagation model can be reduced to the form of Eq.~\eqref{eq:CQGLE2}. Because the local IRS term is a consistent contribution resulting from the general gradient expansion and the bifurcation analysis of non-local PDEs is very involved, we consider the local IRS term in the following. 

Stationary localized solutions of Eq.~\eqref{eq:CQGLE2} can be found using the ansatz $A(t,z)=A(t-v\,z)\,e^{-i\omega z}$, where $\omega$ is the spectral parameter and $v$ is the propagation speed which adds a contribution $(v\,\partial_t+i\omega)\,A$ to the right-hand side of Eq.~\eqref{eq:CQGLE2}. We choose the parameters of Eq.~\eqref{eq:CQGLE2} in a range where soliton explosions exist~\cite{SCAA_PRL2000} and can now track the solutions of the resulting equation in parameter space using the path following technique within the pde2path framework~\cite{uecker2014}. During continuation, both $\omega$ and $v$ become two additional free parameters that are automatically adapted. In order to determine them, we impose additional auxiliary conditions, accounting for the translational and phase-shift symmetries of Eq.~\eqref{eq:CQGLE2} and preventing the continuation algorithm to trivially follow solutions along the corresponding neutral degree of freedom. Note, that in contrast to direct numerical simulations, continuation algorithms are able to track both stable and unstable solutions of the underlying system and make the reconstruction of the whole solution branch, including the information of possible instabilities, feasible. The latter can be followed in the parameter space so that bifurcation diagrams containing the important information about, i.e., a stability region of the studied solution can be created. Note that path continuation techniques are widely employed to obtain different types of solutions of nonlinear ordinary or delay differential equations. However, continuation tools for multidimensional partial differential equations are still scarce. In nonlinear optics, the bifurcation analysis can be quite involved because of the presence of complex fields and different continuous symmetries~\cite{SCAT_OC2001,SCAT_JOSAB2002}. In particular, for the CQGLE~\eqref{eq:CQGLE2} the frequency shift $\omega$ which is connected to the phase-shift invariance is very large, demanding high accuracy calculations: To obtain the LS properly we track $N_t=4 \times 2048$ degrees of freedom (real and imaginary part of the field and their gradients) together with $550$ eigenvalues to resolve the explosion modes.   
 \begin{figure}[h!]
\includegraphics[width=.5\columnwidth]{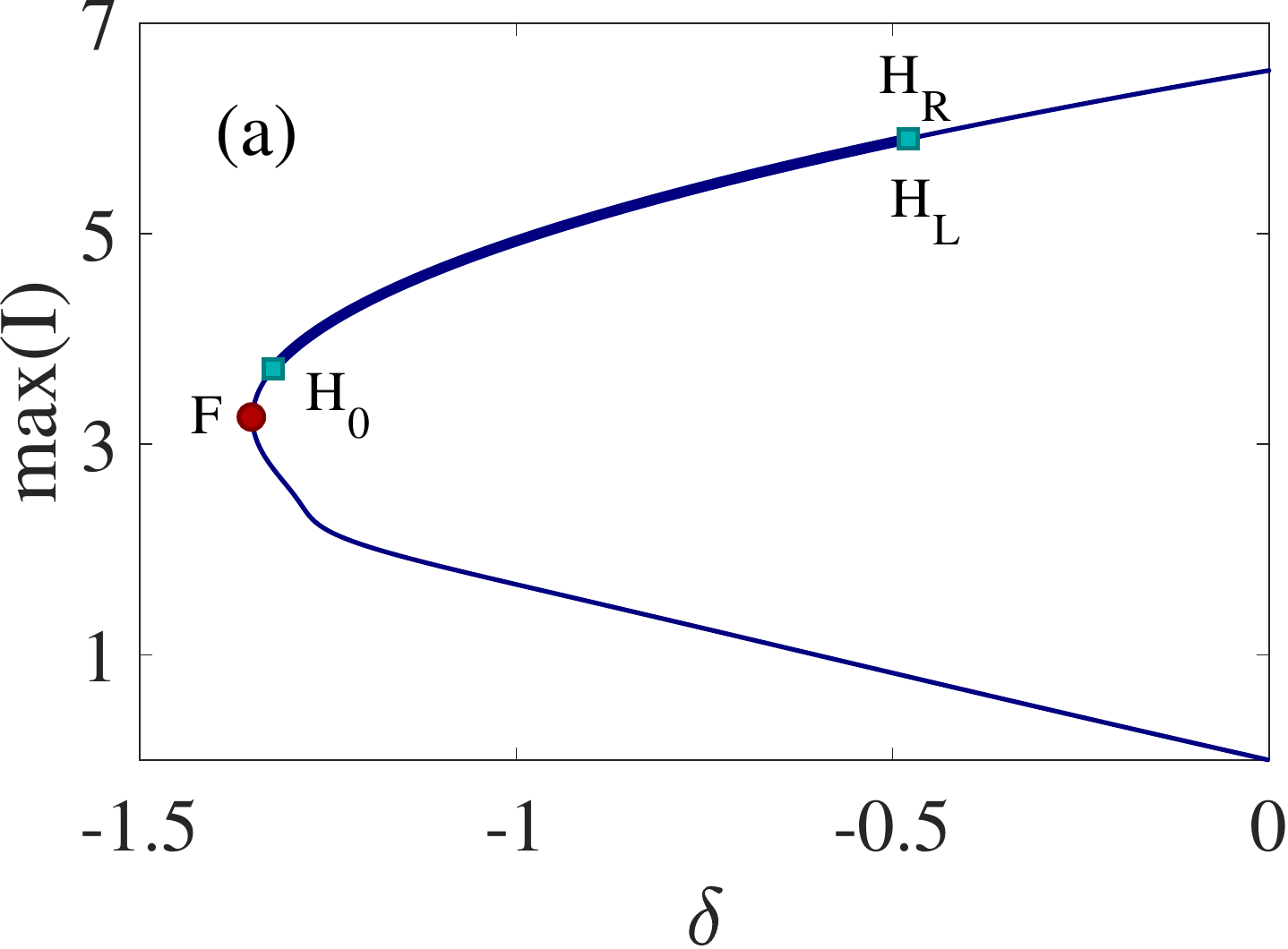}~\includegraphics[width=.48\columnwidth]{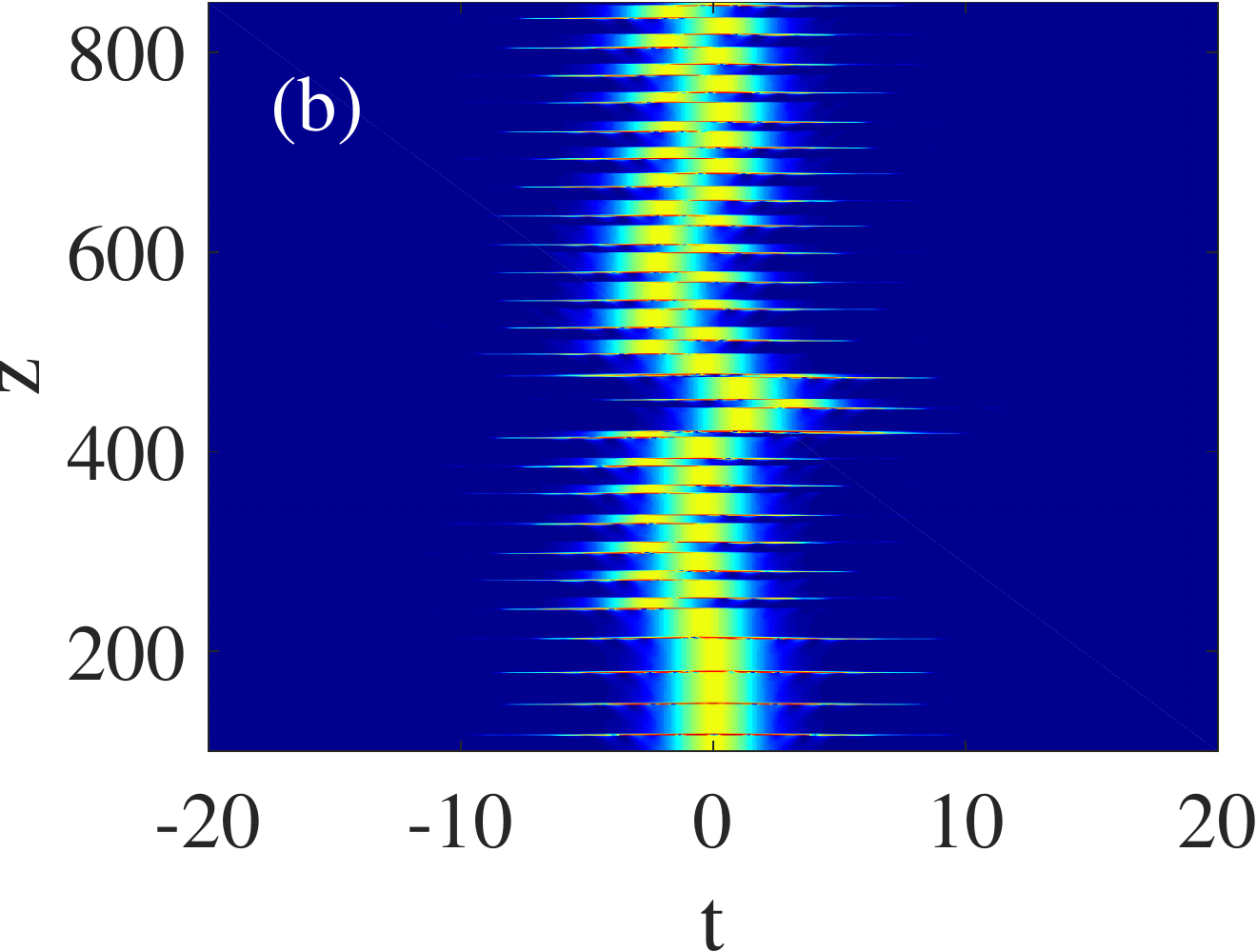}\\
\includegraphics[width=.5\columnwidth]{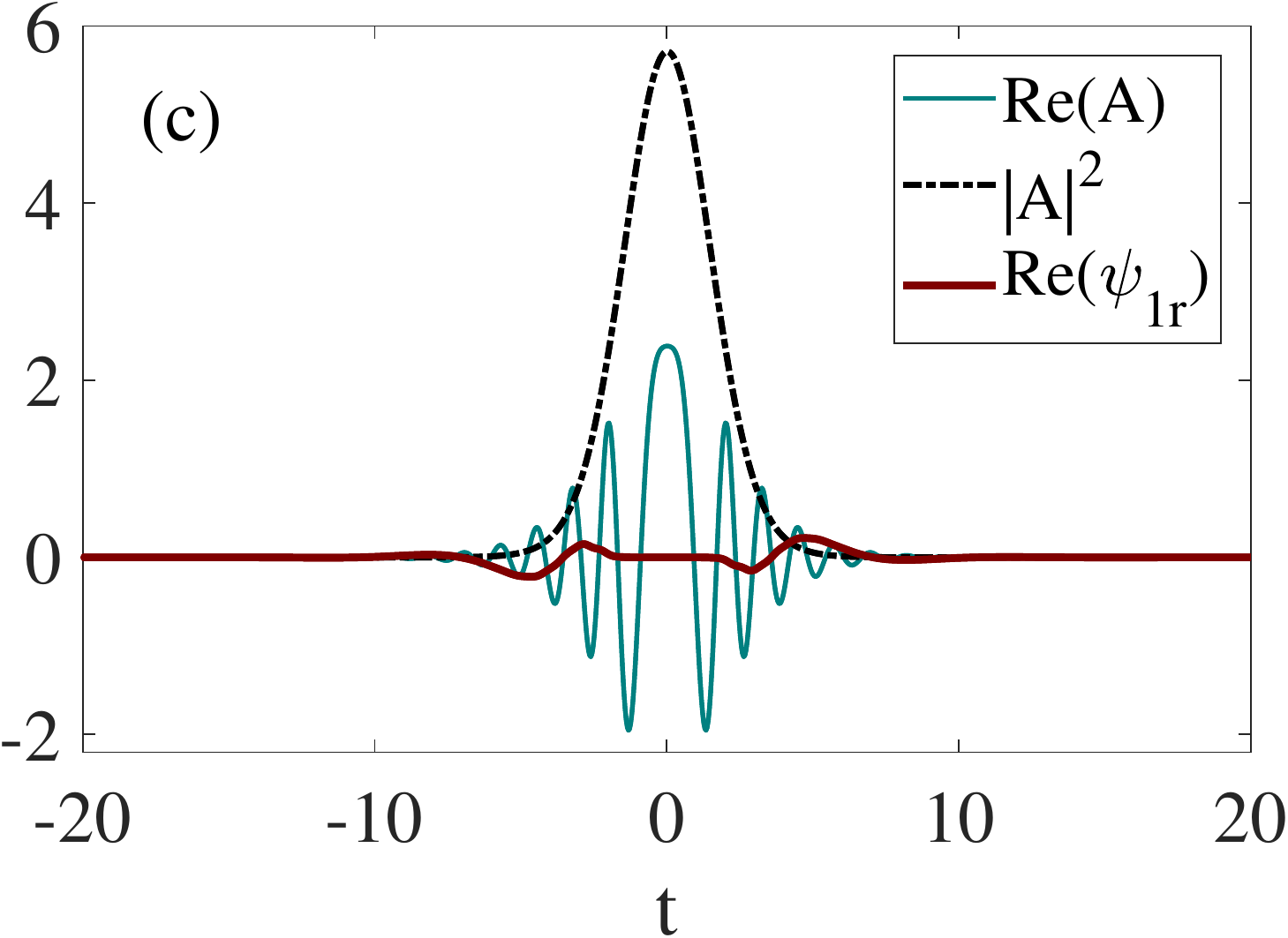}~\includegraphics[width=.5\columnwidth]{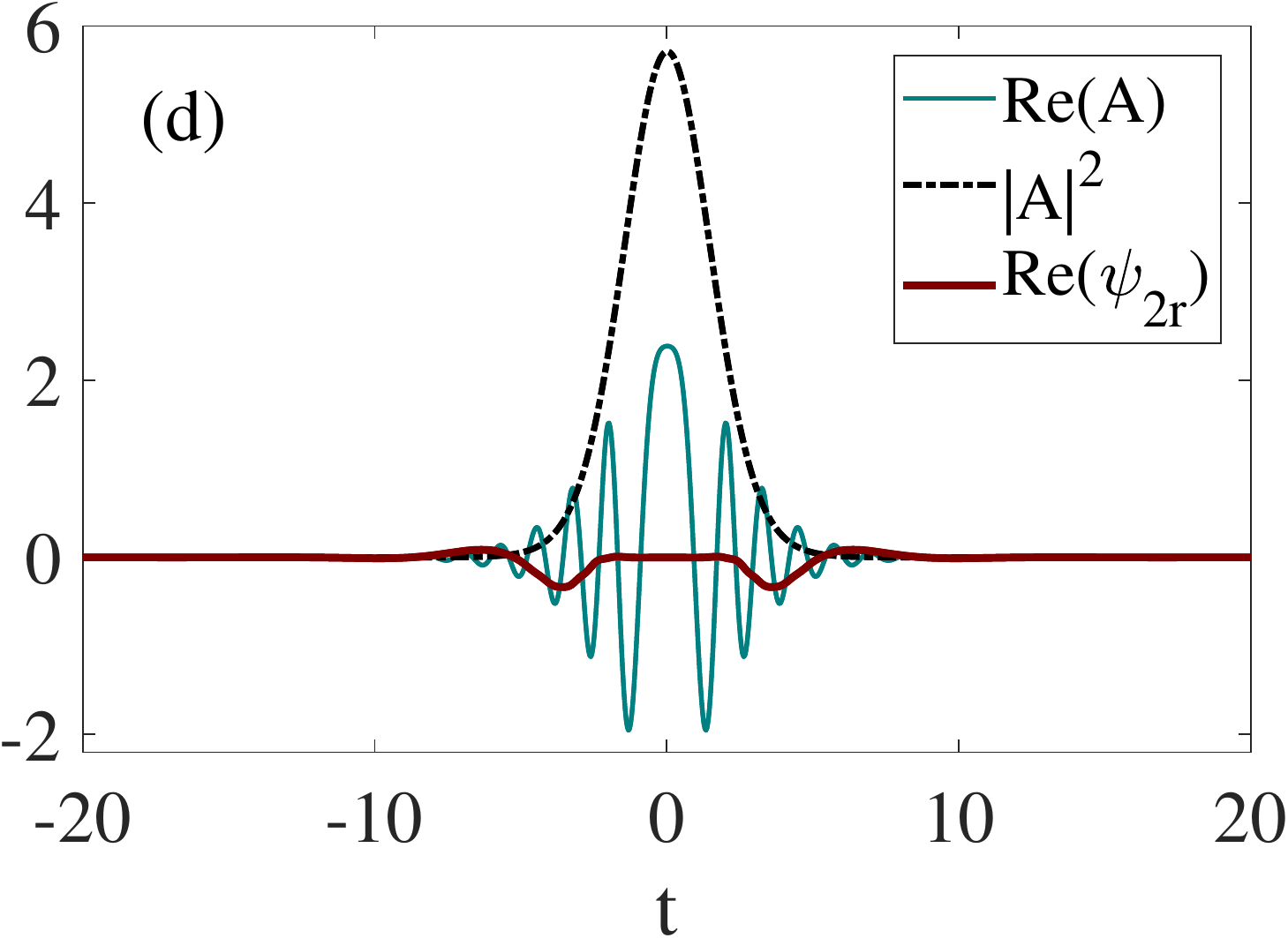}
\caption{(a) A maximal intensity of a single LS of Eq.~\eqref{eq:CQGLE2} as a function of the loss parameter $\delta$ for $\beta_3=s=\tau_r=0$. A LS is stable between a AH point $H_0$ and the threshold of explosions, given by a double AH point $H_{L}, H_R$ at $\delta=-0.5537$. (b) Space time representation of the time evolution of a LS from a DNS for $\delta=-0.3$. (c,d) Real parts of the critical eigenfunctions (red) at $\delta=-0.6$, corresponding to asymmetrical and symmetrical explosions, respectively. Cyan lines represent the $\mathrm{Re}(A)$ field, whereas the black line corresponds to the intensity profile. All quantities are dimensionless. Parameters are: $\epsilon=1.0188$, $\beta=0.125$, $\mu=-0.1$, $\nu=-0.6$.}\label{fig:Branch1d_symm}
\end{figure}

In the absence of HOE terms, the velocity $v$ remains zero and a branch of a stationary LSs emerges when changing the loss parameter $\delta$. Figure~\ref{fig:Branch1d_symm}~(a) shows the maximal intensity $I=|A|^2$ of the field as a function of $\delta$. One can see that a single LS bifurcates subcritically from a homogeneous zero state at $\delta=0$ and experiences a fold $F$ (red point) at some negative value of $\delta$. The LS's stability is governed by the Andronov-Hopf (AH) point $H_0$, where a pair of complex eigenvalues corresponding to the symmetrical  pulsation becomes stable. The LS remains stable for the increasing $\delta$ (thick blue line) until \emph{a double} AH bifurcation point ($H_L,\,H_R$), corresponding to the symmetric (even) and asymmetric (odd) perturbations. The corresponding eigenmode for $\mathrm{Re}(A)$ is shown in cyan together with the intensity profile (black line) in Fig.~\ref{fig:Branch1d_symm}~(c,\,d). One notices that the critical symmetric (asymmetric) modes are localized on both flanks of the LS in phase (antiphase). At the double AH point, two branches of periodic solutions emerge leading to the formation of symmetric and asymmetric explosions at these branches. Figure~\ref{fig:Branch1d_symm}~(b) shows an example of the symmetric-asymmetric exploding LS obtained from a direct numerical simulation (DNS) of Eq.~\eqref{eq:CQGLE2} on the domain of $L_t=100$ with $N_t=1024$ grid points for $\delta=-0.3$ far above from the bifurcation point. Note that in~\cite{ASC_PLA2003} the double AH point and the critical eigenfunctions were found for fixed values of $\delta$ using numerical linear stability analysis. 
%
\begin{figure}[h!]
\includegraphics[width=.5\columnwidth]{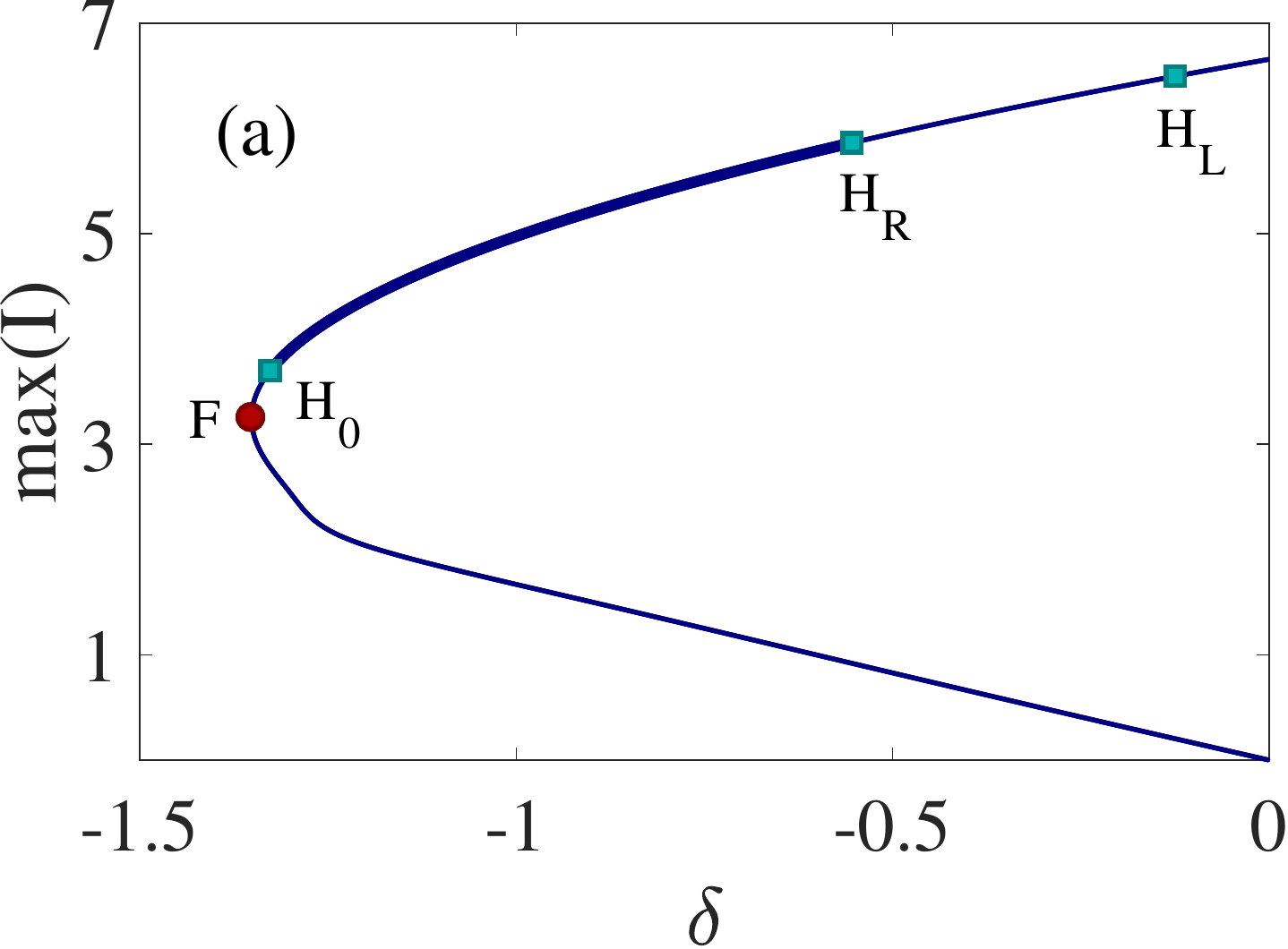}~\includegraphics[width=.5\columnwidth]{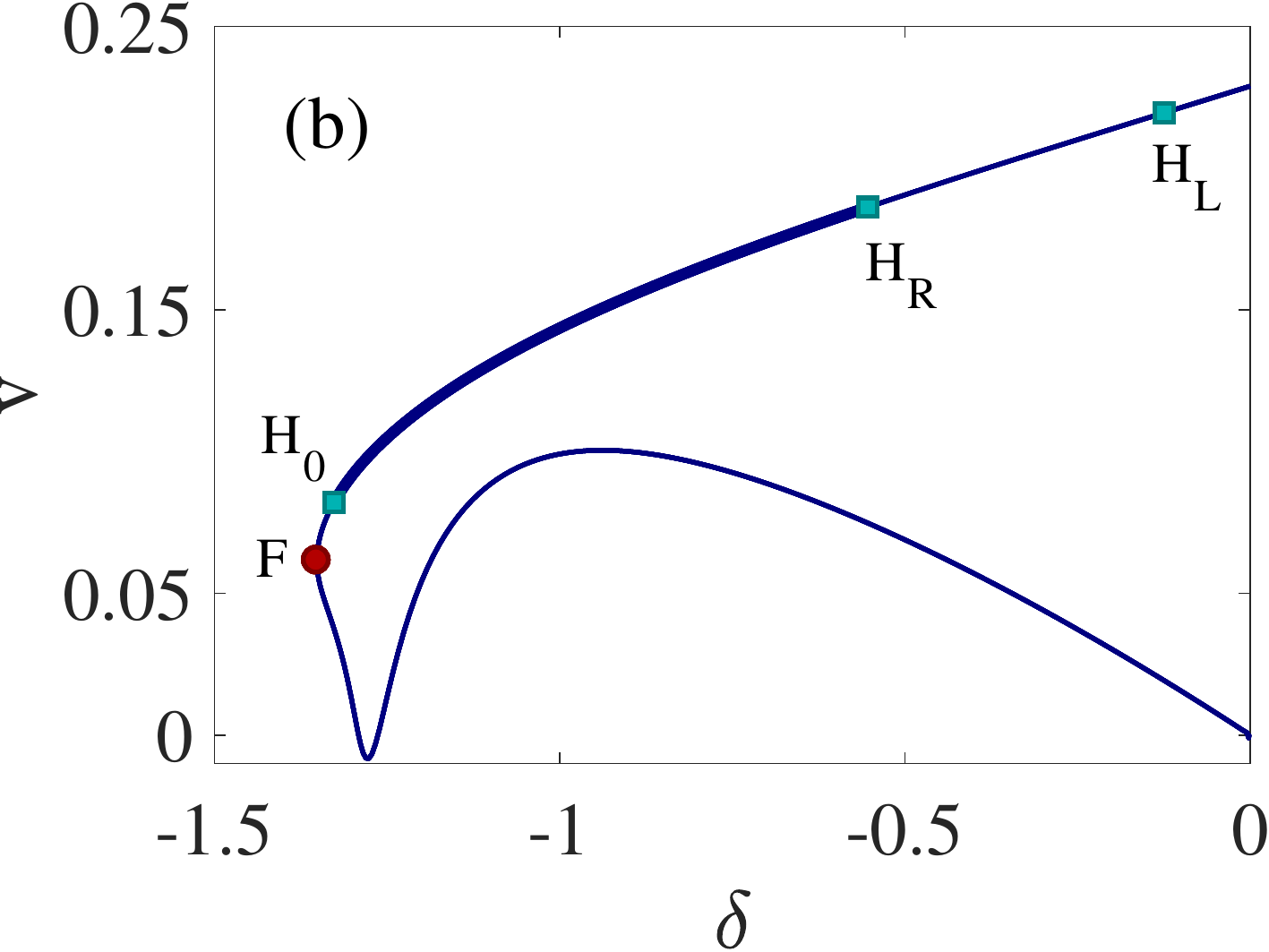}\\
\includegraphics[width=.5\columnwidth]{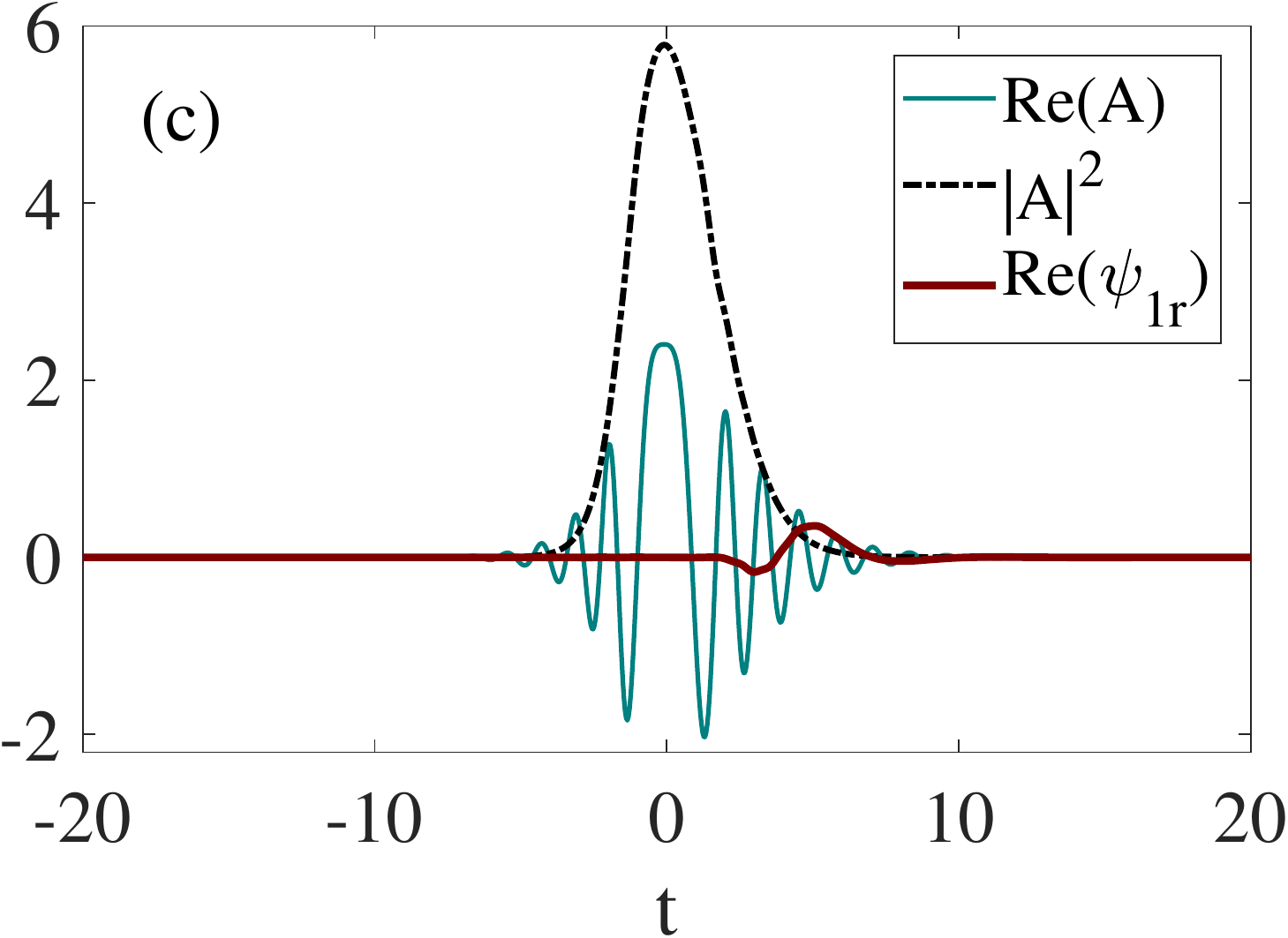}~\includegraphics[width=.5\columnwidth]{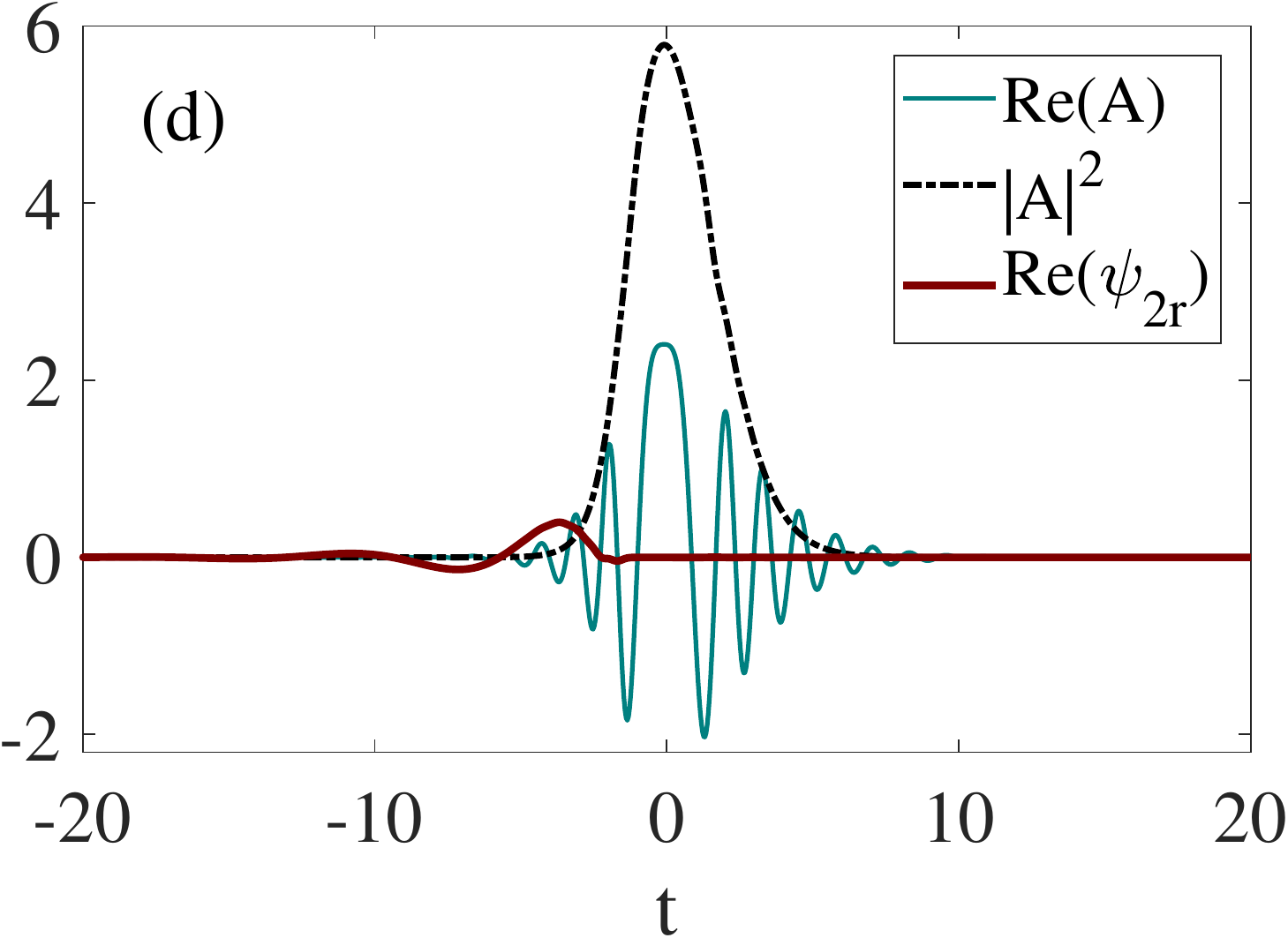}
\caption{ (a,b) Single LS branch of Eq.~\eqref{eq:CQGLE2} as a function $\delta$ for $\beta_3=0.016$, $\tau_R=0.032$, $s=0.009$ where the splitting of AH points $H_R$, $H_L$ can be seen. (a) maximal intensity and (b) drift velocity. A LS is stable between an AH point $H_0$ and the threshold of right-side explosions $H_R$. (c,d) Real parts of the critical eigenfunctions (red), $\mathrm{Re}(A)$ (cyan) and $I=|A|^2$ (black) at $\delta=-0.6$ for the right- and left-side explosions, respectively. All quantities are dimensionless.}\label{fig:Branch1d_tod}
\end{figure}

However, the presence of HOEs significantly changes the behavior of the LS: Each of the TOD, SST and IRS terms in Eq.~\eqref{eq:CQGLE2} break the parity symmetry of the system and affect the leading and falling edges of LSs differently (cf. black dashed lines in Fig.~\ref{fig:Branch1d_tod}~(c,\,d)) and the LSs start to drift as presented in Fig.~\ref{fig:Branch1d_tod}~(b), where the dependence of the LSs velocity $v$ on $\delta$ is shown for non-vanishing fixed values of $\beta_3$, $s$ and $\tau$. One can see, that the unstable part of the branch is a nonlinear function of $\delta$. However, after a fold, the velocity $v$ monotonously increases with $\delta$. In Fig.~\ref{fig:Branch1d_tod}~(a) the peak intensity of a single LS as a function of $\delta$ is presented. Note that the overall shape of the branch remains the same as in the case of vanishing HOEs, namely, a single LS emerges from the homogeneous background at $\delta=0$. The solution branch has a fold at a certain $\delta$ value (cf. the red circle) and gains stability at the AH point $H_0$. However, at the high power branch the double AH point $H_{L,R}$ breaks in the presence of HOEs and two distinct AH points $H_R$ and $H_L$ form. The eigenfunctions corresponding to the perturbation of $\mathrm{Re}(A)$ are not located symmetrically anymore and are localized at left- or right-flanks of the LS as shown in Fig.~\ref{fig:Branch1d_tod}~(c,\,d) (red lines). Again, at both $H_{L,R}$ points, two branches of periodic solutions emerge leading to the formation of left- and right- periodic explosions at these branches (cf. Ref~\cite{CD_PRA2016}). In \cite{FCF_PLA10}, the stationary LSs profiles of Eq.~\eqref{eq:CQGLE2} with a IRS term were reconstructed by using a shooting method. The splitting of $H_{L,R}$ points in the presence of IRS and TOD was shown in \cite{FC_PLA11,CF_PLA2012}. There, a numerical linear stability analysis using Evans functions was used to follow the evolution of critical eigenvalues as a function of IRS and TOD coefficients. In contrast, the continuation algorithms presented here allows to track the eigenvalue spectrum along the whole solution branch giving the complete information about the LS stability.
\begin{figure}[h!]
\includegraphics[width=.5\columnwidth]{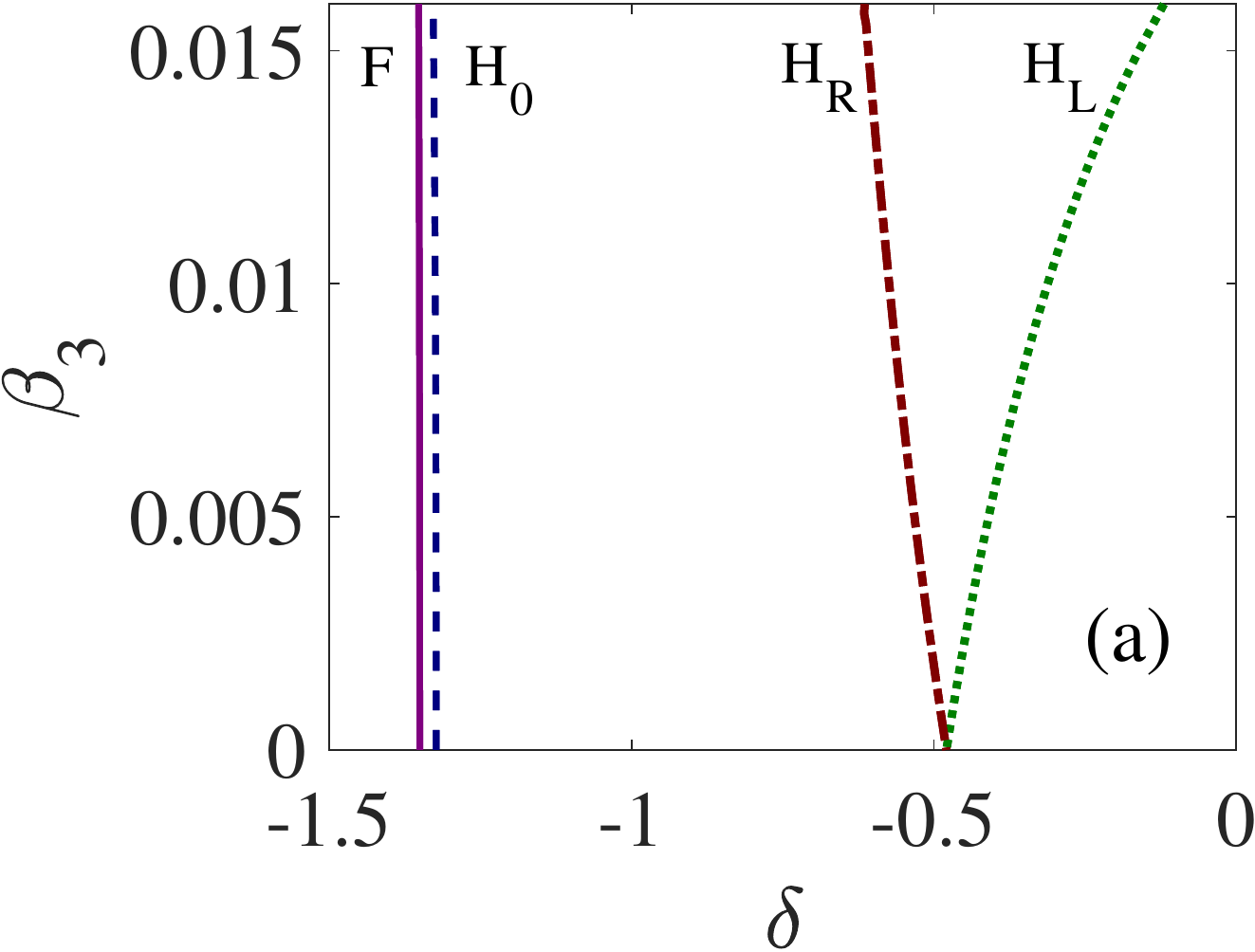}~\includegraphics[width=.5\columnwidth]{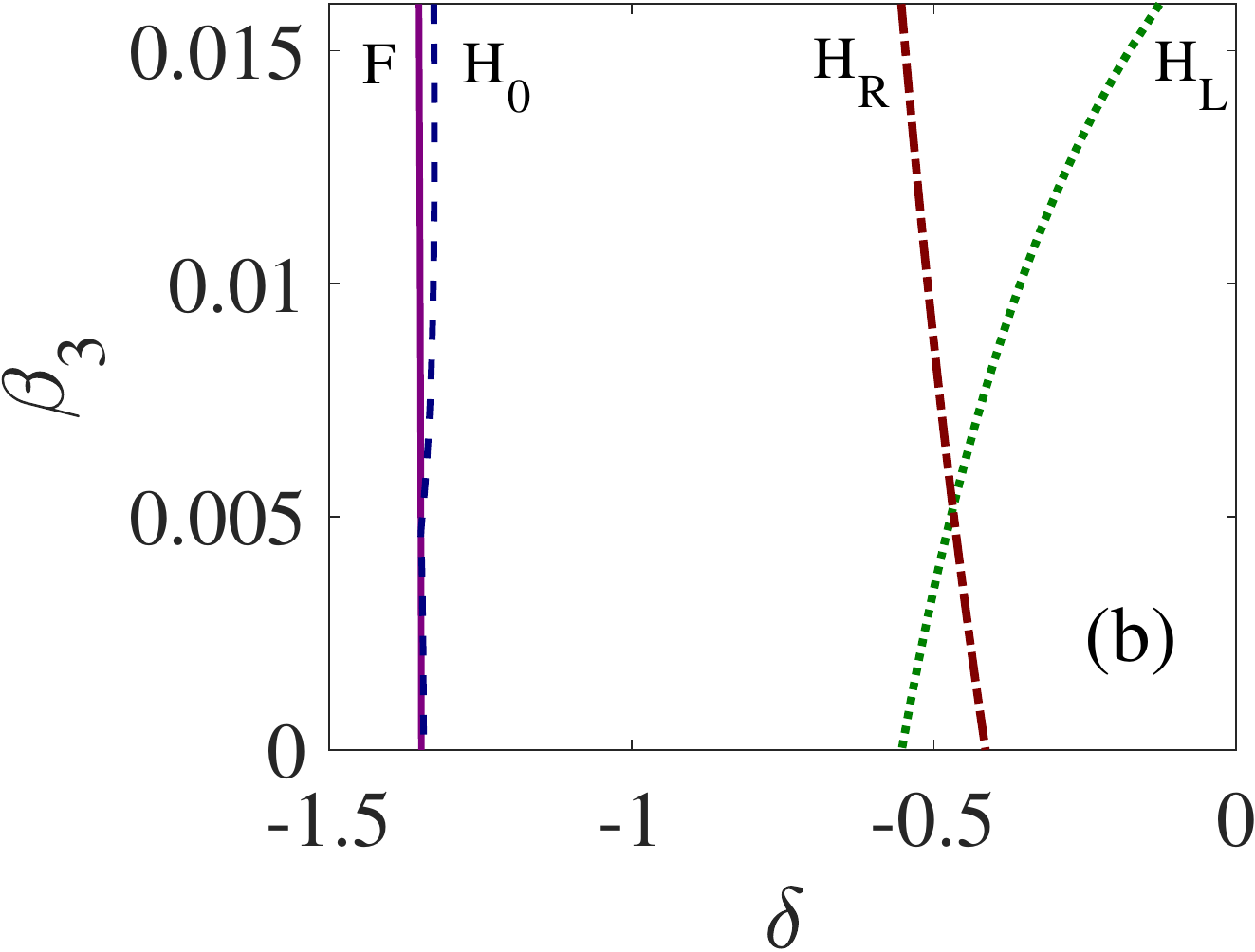}\\
\includegraphics[width=.5\columnwidth]{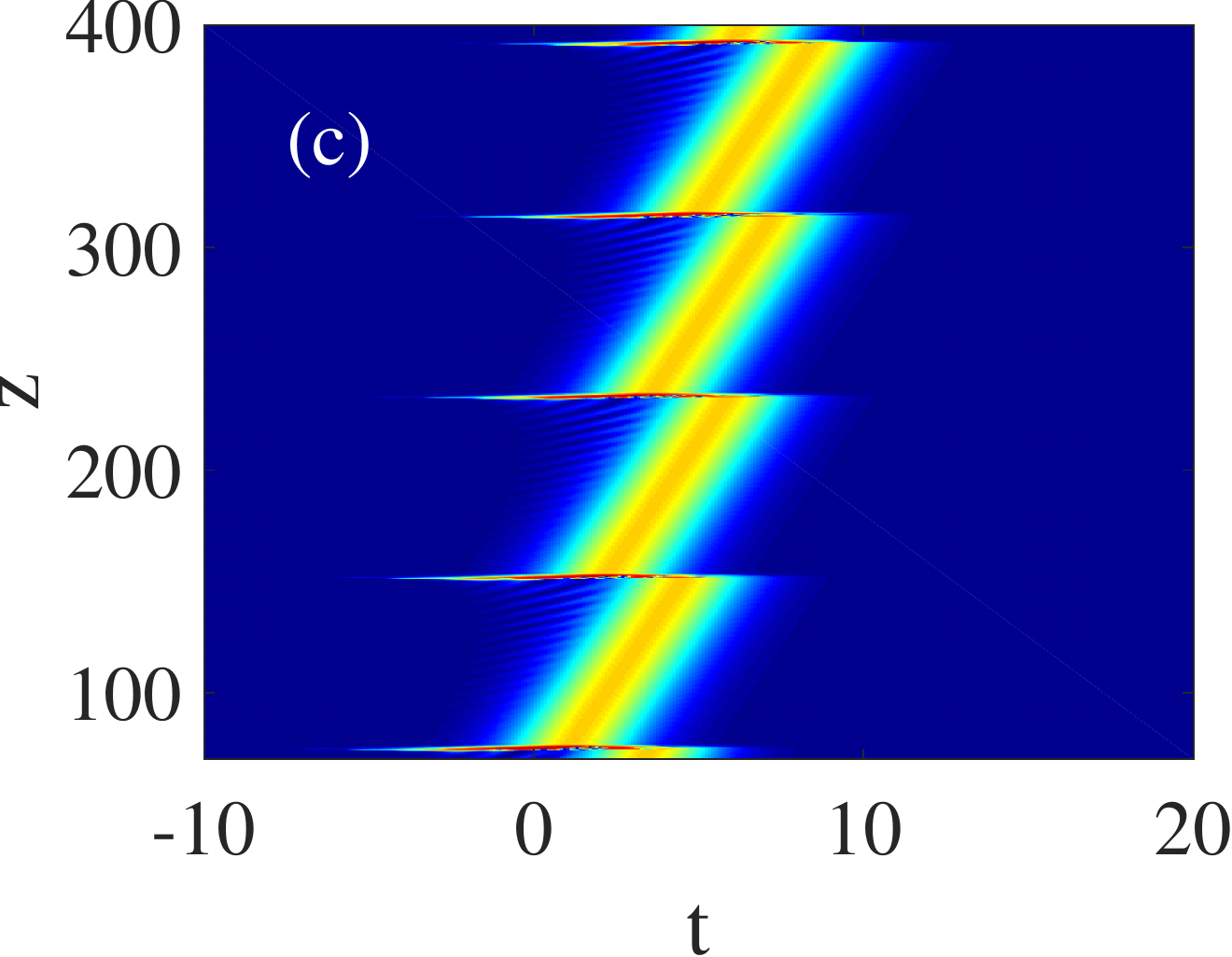}~\includegraphics[width=.5\columnwidth]{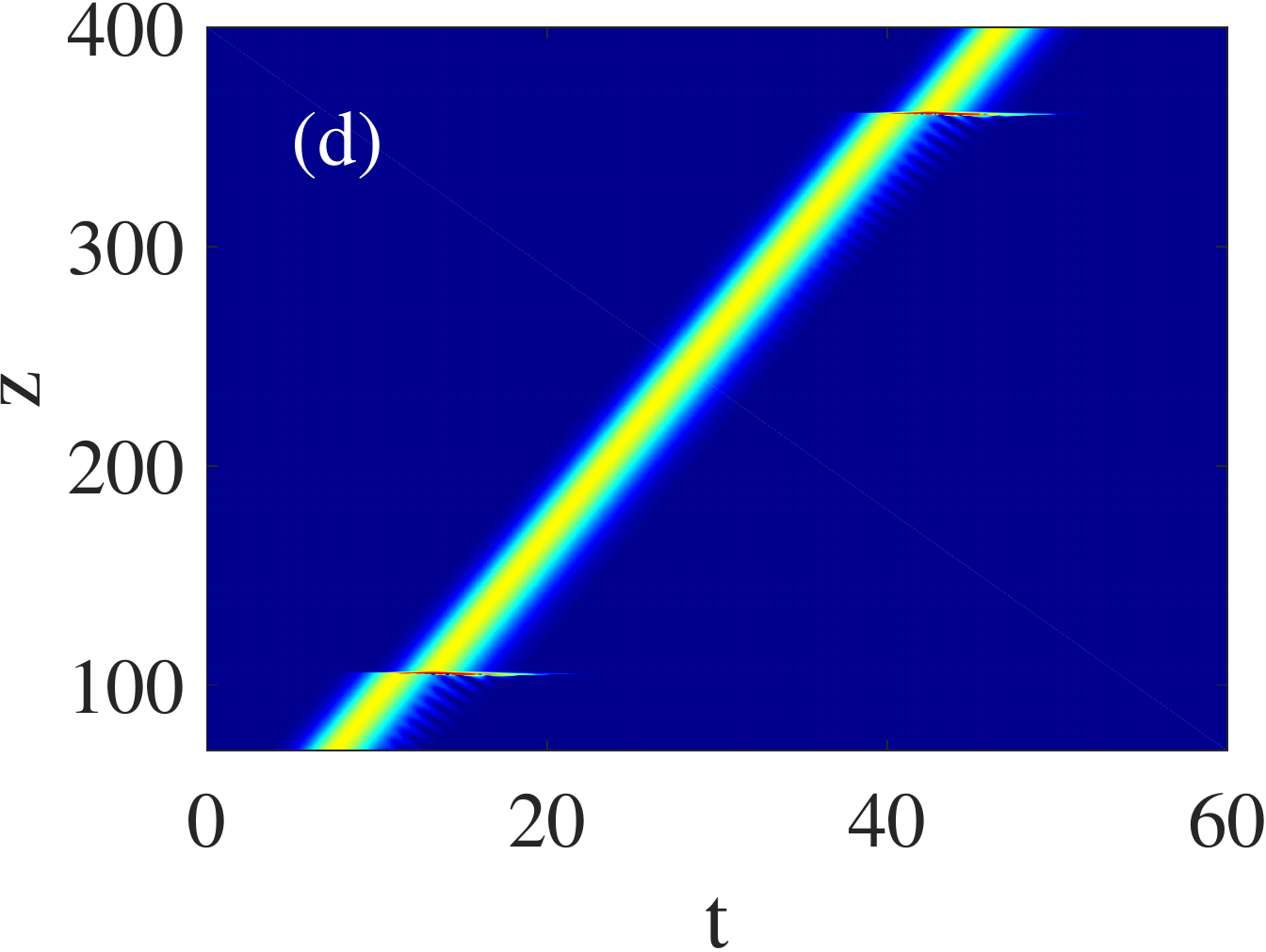}
\caption{(a,b) Bifurcation diagram of a single LS of Eq.~\eqref{eq:CQGLE2} in the $(\delta,\,\beta_3)$ plane showing the splitting of the explosions modes for (a) $\tau_R=0$, $s=0$ and (b) $\tau_R=0.032$, $s=0.009$. A magenta solid line $F$ corresponds to a fold, a blue dashed line $H_0$ is a symmetrical pulsation threshold, whereas $H_L$ and $H_R$ are lines corresponding to the left- and right-side explosions thresholds, respectively. (c,d) Space-time plots showing the (c) left- and (d) right-side explosions calculated within DNS at $(\delta,\,\beta_3)=(-0.45,\,0.001)$ and $(-0.5,\,0.01)$, respectively. All quantities are dimensionless.}\label{fig:BifDiagTOD}
\end{figure}

This splitting of the explosion modes leads to the emergence of two branches of periodic solutions, where periodic one-side left- and right- explosions start to exist. We stress that the splitting of symmetric explosion modes is a general consequence of the presence of HOE terms in the system. It can be interpreted as a result of the breaking of the parity symmetry: That is, any combination of HOEs would lead to the splitting of explosion modes. However, the question if left- or right- explosion mode (or both) is selected is more complicated and the position of the selected left or right modes strongly depends on the amount of HOE coefficients $\beta_3$, $s$ and $\tau$. In order to study the selection of the one-side explosion modes systematically and to determine the region of stability of a LS in the presence of HOEs, we perform a fold and AH point continuation and reconstruct a bifurcation diagram in the plane spanned by $\delta$ and HOE parameters. We start with the case where both IRS and SST effects are absent (i.e. $s=\tau_R=0$) and study the impact of the TOD coefficient $\beta_3$ on the evolution of the fold $F$, as well as the AH points $H_0$, $H_L$ and $H_R$ in the $(\delta,\,\beta_3)$ plane. The results are presented in Fig. ~\ref{fig:BifDiagTOD}~(a). One can see that the position of the fold $F$ as well as of the AH point $H_0$ remain almost unaffected by the TOD. However, even a small amount of positive $\beta_3$ induces a splitting of the double AH point $H_{L,R}$, eliciting first the right-explosions mode $H_R$. That is, for any fixed small $\beta_3>0$, a LS is stable between $H_0$ and $H_R$ lines, and right-side explosions set in first. Increasing $\delta$, the left-side explosion curve $H_L$ can also be crossed and a combination of right- and left- side explosions can be found. Note that a similar behavior and selection order can be achieved by changing the SST coefficient $s$ (or the IRS $\tau_R$) keeping the other two HOEs to zero. Note also that in all these cases the order of the selected explosions modes can be changed by changing the sign of the corresponding HOE term. However, the selection of the $H_L$ or $H_R$ mode can also be tuned by choosing a nonzero amount of all three HOE coefficients as shown in Fig.~\ref{fig:BifDiagTOD}~(b), where the bifurcation diagram in the $(\delta,\,\beta_3)$ plane is presented for non-vanishing values of $\tau_R$ and $s$. One can see that as in the case of zero SST and IRS terms, the positions of the fold $F$ and the AH point $H_0$ remain almost the same in $\delta$ and the double AH point splits again. However, a presence of nonzero $s$ and $\tau_R$ shifts the $H_L$, $H_R$ crossing point in the direction of the positive TOD coefficient $\beta_3$. This leads to the formation of two regions in the parameter space: Whereas after the crossing point, the right-side explosions are selected (red dashed line), a new region emerges, where the left-side explosion mode wins for small values of $\beta_3$ (green dotted line). That is, depending on the loss parameter $\delta$ one can select left- or right-side explosions depending on the values of the TOD. Two examples of the direct numerical simulations of Eq.~\eqref{eq:CQGLE2} showing the left- and right-side periodic explosions of a single LS are presented in Fig.~\ref{fig:BifDiagTOD}~(c, d). Note that in this case both left- and right-side explosions keep the positive propagation direction. 

\begin{figure}[h!]
\includegraphics[width=.5\columnwidth]{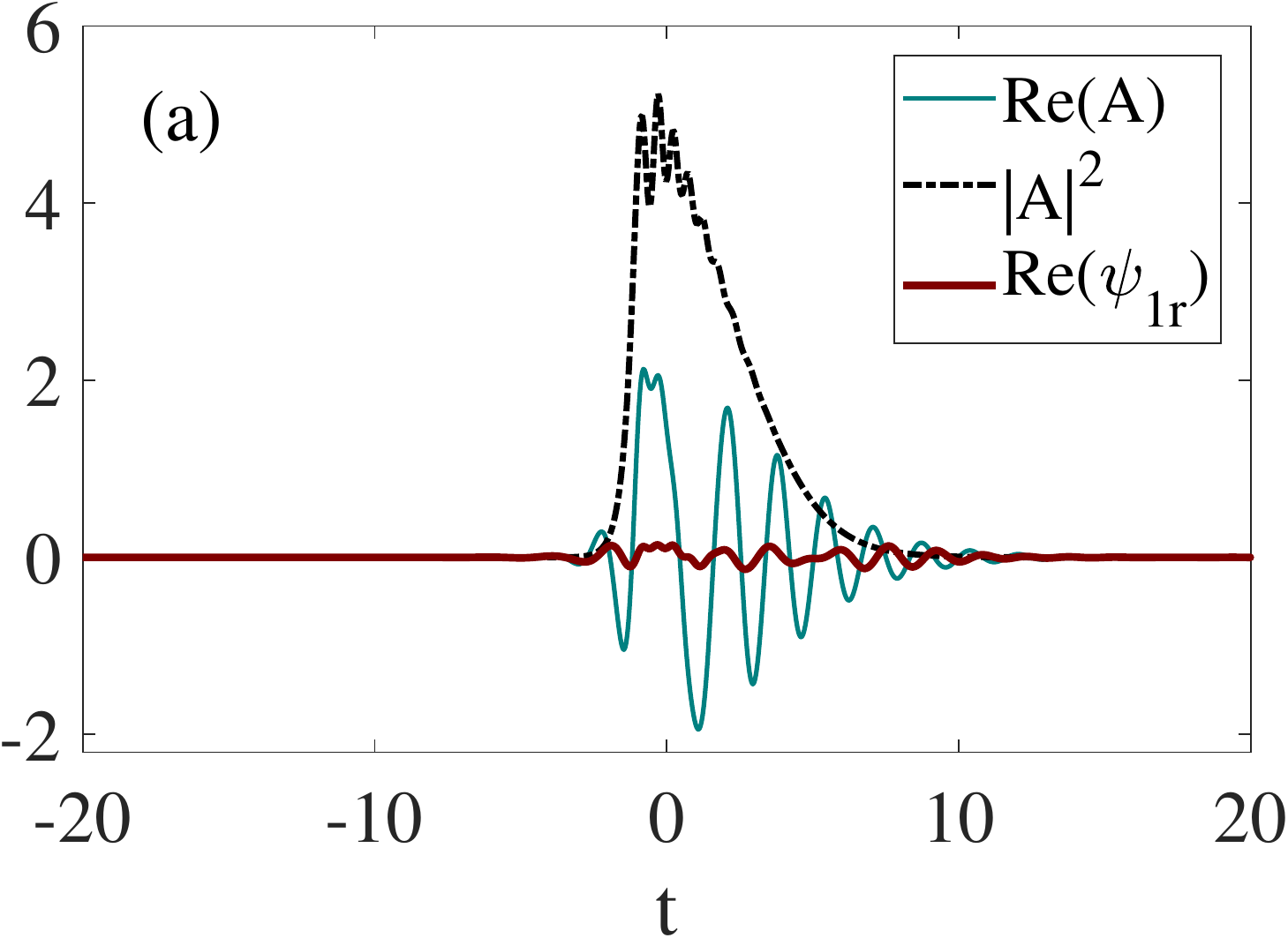}~\includegraphics[width=.5\columnwidth]{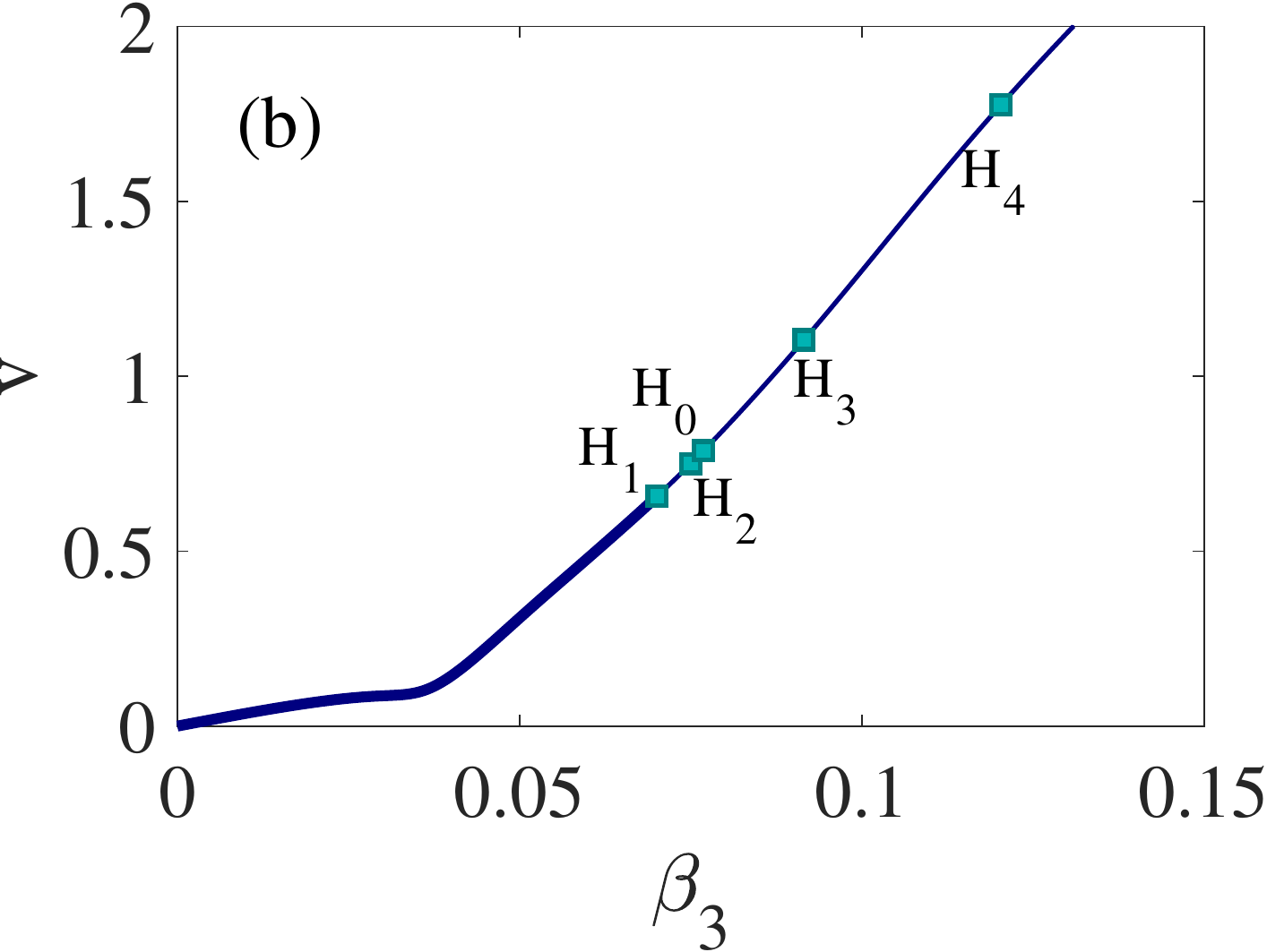} \\
\includegraphics[width=.5\columnwidth]{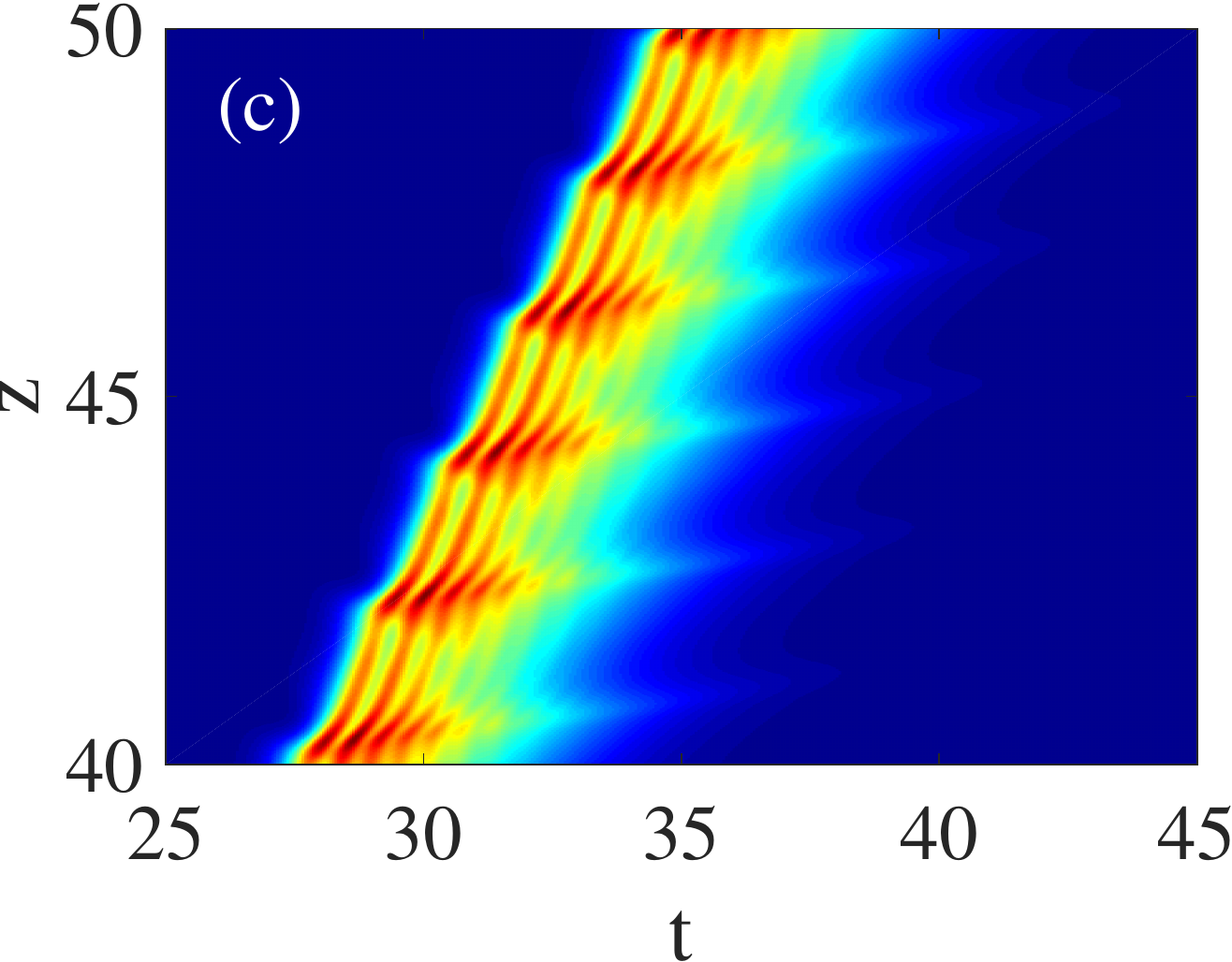}~\includegraphics[width=.5\columnwidth]{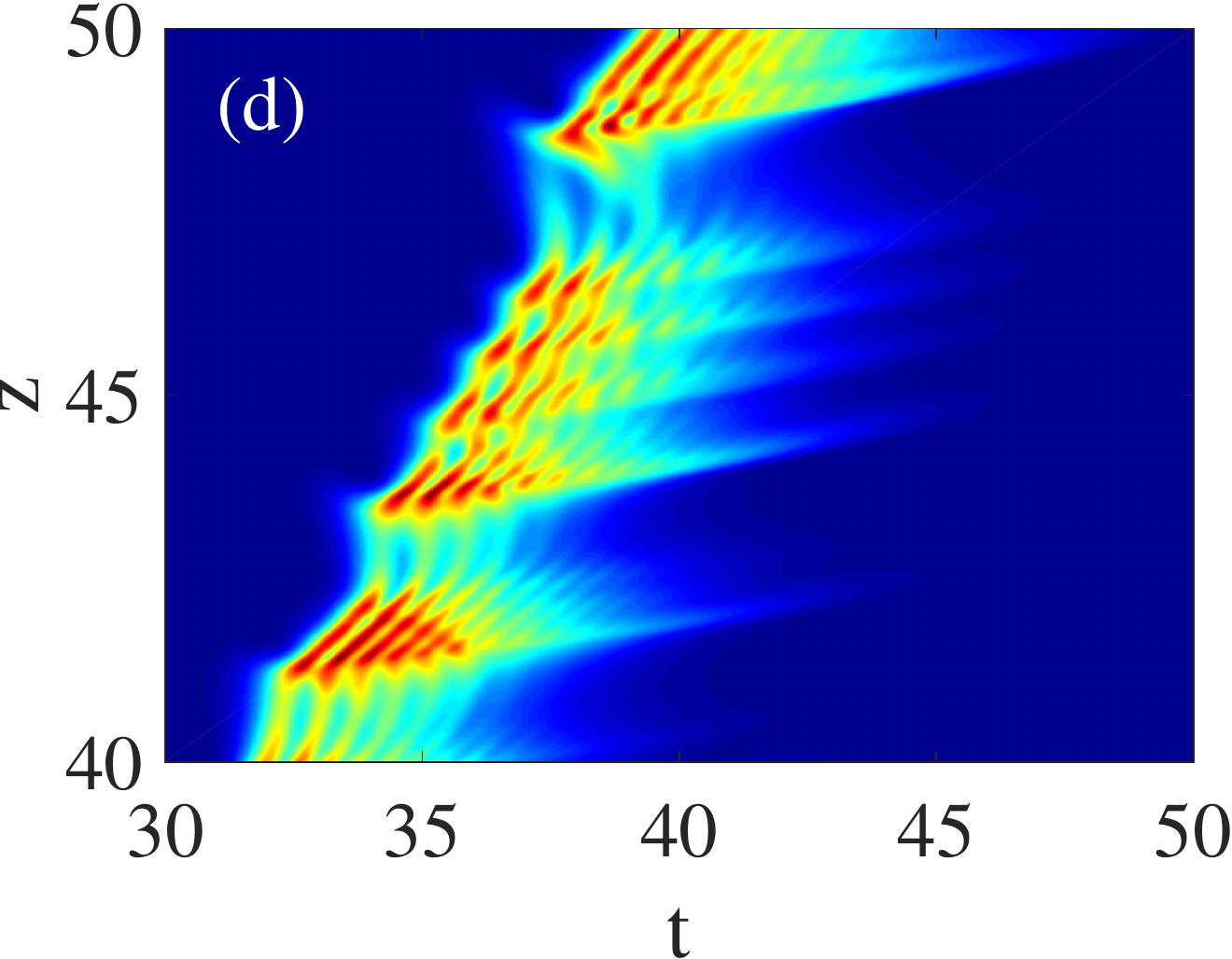}\\
\caption{(a) Intensity profile (black dashed line) of a single LS of Eq.~\eqref{eq:CQGLE2} calculated at $\beta_3=0.068$ and $\delta=-1$ together with the real part of the critical eigenfunction $H_1$ (red) and $\mathrm{Re}(A)$ (cyan). (b) Drift velocity $v$ as a function of the TOD coefficient $\beta_3$. The points $H_{1-4}$ correspond to the TOD induced AH bifurcations, whereas $H_0$ corresponds to a symmetric pulsation (cf. Fig. \ref{fig:BifDiagTOD}). (c,d) Time evolution of the LS calculated from DNS for $\beta_3=0.07$ and $\beta_3=0.08$, respectively. All quantities are dimensionless.}\label{fig:TODInduced}
\end{figure}

Figure~\ref{fig:BifDiagTOD} indicates that the presence of HOEs does not change the number of critical modes responsible for the primary destabilization of the LS but only the explosion modes. However, this situation changes dramatically when, e.g., the TOD coefficient $\beta_3$ is increased further. To study the impact of TOD on the dynamics of the LS, we first fixed both SST and IRS terms to zero and performed a continuation in $\beta_3$, keeping the value of the loss parameter $\delta$ fixed at the value inside the stability region. As was mentioned above, TOD affects the leading and falling edges of a LS differently and leads to the emergence of an asymmetric pulse whose right (trailing) tail contains a weakly decaying dispersive wave which is visible only in logarithmic scale if $\beta_3$ is very small. However, with increase of $\beta_3$, the effect of TOD becomes visible and a LS with a strongly oscillating profile emerges as shown in Fig.~\ref{fig:TODInduced}~(a). This kind of LS solution can be stable for a large range of $\beta_3$ and finally looses its stability in a new, TOD-induced, AH bifurcation $H_1$. Our results are depicted in Fig.~\ref{fig:TODInduced}~(b), where the velocity of a LS is shown as a function of $\beta_3$. The corresponding critical eigenfunction is depicted in red in Fig.~\ref{fig:TODInduced}~(a) together with $\mathrm{Re}(A)$ whereas in the panel (c) a time evolution of the LS above the bifurcation point is shown.  One can see, that after the $H_1$ bifurcation point, a  number of other AH modes become unstable, leading to a complex chaotic dynamics of the pulse (see Fig.~\ref{fig:TODInduced}~(d)). This scenario is different from the one responsible for the explosions: They are formed at the periodic branch emerging form $H_{L,R}$ points, whereas the periodic and chaotic behavior here result from the chain of the unstable AH modes acting on the LSs. The full bifurcation analysis in the region of high $\beta_3$ is very involved and is beyond the scope of this paper. However, our preliminary results indicate that the mode $H_1$ remains the most unstable mode for the large region of $\delta$ where the explosion modes are not responsible for the primary instabilities. It is overtaken by the $H_L$ mode only if the gain value is increased to a value comparable with those shown in Fig.~\ref{fig:BifDiagTOD}. Furthermore, for non-vanishing $s$ and $\tau_R$ the overall behavior remains qualitatively identical and the position of the first TOD induced instability shifts to smaller $\beta_3$ values, making the region of the LS stability even smaller. This result is markedly different from the conclusion obtained for the Lugiato-Lefever equation with TOD, where a stabilization of a LS by TOD was demonstrated in a wide range of parameters~\cite{PGL-OL-14}.

In conclusion, we studied the impact of self-frequency shift, self-steepening, and third-order dispersion on the selection mechanism of soliton explosions in the CQGLE. Using path continuation techniques, we showed how the interplay between different HOEs inducing the splitting of the symmetric and asymmetric explosion modes results in the controllable selection of left- and right- one-side periodic explosions. Finally we found that the third-order dispersion leads to new pulsating regular and irregular instabilities, which leads to the significant reduction of the stability region of the single LS.

\section*{Funding information} 
\noindent MINECO Project COMBINA (TEC2015-65212-C3-3-P). 

\section*{Acknowledgments}
S.G. acknowledges the Universitat de les Illes Balears
for funding a stay where part of this work was developed.


\end{document}